\documentclass[prd,preprint,superscriptaddress,amsmath,amssymb,nofootinbib]{revtex4}
\usepackage{graphicx}
\usepackage{dcolumn}
\usepackage{bm}
\usepackage{amssymb}
\usepackage{amsmath}
\usepackage{epsfig}    
\usepackage{color}
\usepackage{slashed}
\usepackage{hhline}
\usepackage{bbm}
\usepackage{tikz-feynman}
\tikzfeynmanset{compat=1.1.0}

\def\be{\begin{equation}}
\def\ee{\end{equation}}
\newcommand{\bea}{\begin{eqnarray}}
\newcommand{\eea}{\end{eqnarray}}

\begin{document}

\title{A natural realization of inverse seesaw model in a non-invertible selection rule}

\author{Shilpa Jangid }
\email{shilpajangid123@gmail.com}
\affiliation{Shiv Nadar IoE Deemed to be University, Gautam Buddha Nagar, Uttar Pradesh, 201314, India}

\author{ Hiroshi Okada}
\email{hiroshi3okada@htu.edu.cn}
\affiliation{Department of Physics, Henan Normal University, Xinxiang 453007, China}

\date{\today}

\begin{abstract}
We propose a natural realization of neutral fermion hierarchies within the inverse seesaw framework by imposing a $Z_3$ Tambara–Yamagami selection rule, which is a minimal symmetry to realize our model.
 The selection rule forbids Majorana mass terms for $N_R$ and $N_L$ at tree level,
while these terms are generated at one-loop through the radiative symmetry breaking of this selection rule.
 To realize such loop corrections, we introduce additional neutral boson and fermion fields, one of which can serve as a viable dark matter candidate. Finally, we present best-fit results for the neutrino sector in both normal and inverted hierarchies, consistent with recent experimental data.

 \end{abstract}
\maketitle

\section{Introduction}
Inverse Seesaw (ISS) is a verifiable scenario of generating tiny neutrino masses within TeV scale~\cite{Mohapatra:1986bd, Wyler:1982dd}.
The model requires two types of neutral fermions $N_R$ and $N_L$, where $N_R$ has right-handed chirality and $N_L$ has left-handed chirality.
Then, we generally have the following mass terms in collaboration with the standard model (SM) particles, 
\begin{align}
m_{D_R} {\overline N_R} \nu_L +m_{D_L} {\overline N_L^C} \nu_L +
M_D {\overline N_R}N_L + m_R {\overline N_R} N_R^C +m_L {\overline N_L^C} N_L +{\rm c.c.}\ , \label{eq:iss}
\end{align}
where $m_{D_R}\equiv y_{D_R} v_H/\sqrt2$ and $m_{D_L}\equiv y_{D_L} v_H/\sqrt2$ are respectively obtained after spontaneous symmetry breaking, $y_{D_R}{\overline N_R} H L_L$ and $y_{D_L}{\overline N_L^C} H  L_L$.
$H$ is the SM Higgs boson whose vacuum expectation value (VEV) is denoted by $\langle H\rangle\equiv [0,v_H/\sqrt2]^T$ and the left-handed active neutrino $\nu_L$ is a component of $L_L\equiv [\nu_L,\ell_L]^T$ in the SM lepton doublet field.
Then, the mass matrix for the neutral fermions in basis of $[\nu_L,N^C_R,N_L]^T$ is given by
\begin{align}
\begin{pmatrix}
 0 & m_{D_R}^T & m_{D_L}^T  \\
 m_{D_R} & m_R & M_D \\
 m_{D_L} & M_D^T & m_L 
  \end{pmatrix}.
\end{align}
When we impose the following mass hierarchies among them~\footnote{There are several ways of the hierarchies to realize the inverse seesaw. }
\begin{align}
 m_{D_L} \ll
   m_R  , m_L \ll m_{D_R} \lesssim  M_D, \label{eq:mass-order}
\end{align}
the active neutrino mass matrix is found as
\begin{align}
m_\nu\simeq m^T_{D_R} (M_D^T)^{-1} m_L M_D^{-1} m_{D_R}. \label{eq:iss}
\end{align}
%
Since the active neutrinos mix with the heavier neutral fermion sector, we need to consider constraints from non-unitarity.
The constraints can be characterized by the hermitian matrix $(F\equiv)  -M_D^{-1} m_{D_R}$, and
the non-unitarity matrix, which is denoted by $U'$, is given by the following form 
\begin{align}
U'\equiv \left(1-\frac12 FF^\dag\right) U.
\end{align}
$U$ is a unitary matrix to diagonalize the active neutrino mass matrix $m_\nu$. Moreover, $U$ is a testable matrix that encodes several observables such as mixing angles and Dirac CP phase under the diagonal mass matrix for charged-leptons. In fact, we work on this setup in our main discussion below. 
%
Considering several experimental results such as the effective Weinberg angle, SM $W$ boson mass, several ratios of $Z$ boson fermionic decays, invisible decay of $Z$, electroweak universality, measurements of the quark mixing matrix, and lepton flavor violations~\cite{Fernandez-Martinez:2016lgt},
the most stringent constraint comes from 1-2 component of $|FF^\dag|$~\cite{Agostinho:2017wfs, Das:2017ski} and  is found as
\begin{align}
|F| \lesssim 4.90\times 10^{-3}. \label{eq:f}
\end{align}
It suggests that $m_{D_R}\ll M_D$ is experimentally assured in Eq.~(\ref{eq:mass-order}).
In a theoretical viewpoint, $M_D$ can be expected to originate from new scale. When we suppose our new physics is within TeV scale,
$M_D$ would be ${\cal O}(1)$ TeV. On the other hand, $m_{D_R}(\sim v_H y_{D_R})$ is generated after the spontaneous electroweak symmetry breaking whose scale is $v_H\sim {\cal O}$(100) GeV.
Thus, if the Yukawa coupling is $y_{D_R}\sim {\cal O}(0.01)$, the required hierarchy in Eq.~(\ref{eq:mass-order}) can be achieved.
However, we need to introduce an additional symmetry or new particles theoretically to understand the following hierarchies
\begin{align} m_{D_L} \ll  m_R  , m_L \ll m_{D_R}.  \label{eq:mass-order-iss} \end{align}
%
For examples, refs~\cite{ Nomura:2018cfu,  Dey:2019cts, Nomura:2020cog} show that $m_{R,L}$ are generated via VEV of multiplet boson  that is restricted by the $\rho$ parameter and its VEV has to be less than ${\cal O}(1)$ GeV while $m_{D_L}$ is forbidden by an additional symmetry. 
Thus $m_{R,L}\ll m_{D_{R}}$ is theoretically explained.
On the other hand, refs.~\cite{Nomura:2018ktz, Nomura:2024jxc, Nomura:2021adf, Okada:2012np} demonstrate that $m_{R,L}$ are induced via one-loop level while forbidding $m_{D_L}$ via an additional symmetry. Hence, $m_{R,L}\ll m_{D_{R}}$ is explained well.

In this paper, we provide a natural explanation for these mass hierarchies in Eq.~(\ref{eq:mass-order-iss}) by invoking a novel non-invertible symmetry: $Z_3$ Tambara-Yamagami ($Z_3$-TY) selection rule, which is recently applied to phenomenologies~\cite{Kobayashi:2025cwx}.
\footnote{Several studies have applied these kinds of symmetries to phenomenologies~\cite{Okada:2025kfm, Choi:2022jqy,Cordova:2022fhg,Cordova:2022ieu,Cordova:2024ypu,Kobayashi:2024cvp,Kobayashi:2024yqq,Kobayashi:2025znw,Suzuki:2025oov,Liang:2025dkm,Kobayashi:2025ldi,Kobayashi:2025lar,Kobayashi:2025cwx,Nomura:2025sod,Dong:2025jra,Nomura:2025yoa,Chen:2025awz, Kobayashi:2025thd, Suzuki:2025bxg}.}
{
The $Z_3$-TY selection rule provides a compact non-invertible framework for realizing the required operator structure of our model. In particular, our construction relies on two important fusion properties:
$n\otimes n\in{\mathbbm I}$ and $a\otimes  a \notin{\mathbbm I}$,
where \(a\) and \(a^*\) denote the invertible simple objects, and \(n\) denotes the non-invertible simple object.
The first property allows a bilinear composed of two fields carrying charge $n$, such as a Majorana mass term for a fermion or a quadratic mass term for a real scalar. The second property forbids a bilinear composed of two fields carrying charge $a$, while \(a\otimes a^*={\mathbbm I}\) allows the corresponding conjugate-field bilinear, such as \(|\phi_a|^2\) for a complex scalar.
}
%
%
%
All the $Z_3$-TY multiplication rules are given by~\cite{Dong:2025jra, Chang:2018iay}
\begin{align}
a\otimes a =a^*,\quad a^*\otimes a^* =a,\quad a\otimes a^*={\mathbbm I},\quad a\otimes  n=a^*\otimes n=n,\quad n\otimes  n={\mathbbm I}\oplus a\oplus a^*,
\end{align}
where $a^{(*)}$ and $n$ are commutable each other.
This symmetry possesses several favorable features for model buildings, especially the radiative symmetry breaking mechanism~\cite{Kobayashi:2025cwx}. 
The $Z_3$-TY  is
strategically employed to forbid tree-level Majorana mass terms $m_L$ and $m_R$ for the singlet fermions $N_L$ and $N_R$ assigned by $a$.
Although $m_{D_L}$ is forbidden by this symmetry at all the loop-orders,
$m_L$ and $m_R$ are generated radiatively at the one-loop level through the dark matter (DM) candidates assigned by $n$ to accommodate the Majorana feature.
The DM candidates are not just additional particles, but essential components for the radiative generation of the ISS.
$m_L$ and $m_R$ break the $Z_3$-TY symmetry as can be seen in Fig.~\ref{fig:Mass}. We will explain the details in the next section.
This mechanism yields the small mass scales required by the inverse seesaw framework without necessitating fine-tuned parameters.~\footnote{In fact, one can straightforwardly derive Ma model~\cite{Ma:2006km} via this mechanism. It is currently known that Ma model of this mechanism can be constructed by only two selection rules; $Z_3$-TY and $Z_3$ gauging of $\mathbbm{Z}_3\times \mathbbm{Z'}_3$~\cite{Dong:2025jra}.}
{
Such a radiative hierarchy cannot be realized through a simple Abelian charge assignment, as we demonstrate in Sec. II.B.}
%

This paper is organized as follows. In Section~II, we present the model with mass spectrum, relevant interactions and formulas for phenomenology of our interest.
In Section III, we perform numerical analysis to search for allowed parameter region and predicted flavor observables. 
Finally, we devote Section~IV to the summary and conclusion.

\begin{figure}[t]
\begin{center}
\begin{tikzpicture}
\begin{feynman}[large]
\vertex (a1) {\(\bar N_{L/R}^C(a)\)};
\vertex [right=1.8cm of a1] (b1);
\vertex [right=4.0cm of b1] (c1);
\vertex [right=1.5cm of c1] (d1) {\(N_{L/R}(a)\)};
\vertex [right=2.0cm of b1] (m1);
\vertex [above=2.0cm of m1] (e1);
\diagram [medium] {
(a1) -- (b1) -- [insertion={[size=3pt]0.5}, edge label'=\(X_R(n)~~~~~~~~~~~~~X_R(n)\)] (c1) -- (d1),
(b1) -- [scalar, quarter left, edge label=\(S(n)\)] (e1) -- [scalar, quarter left, edge label=\(S(n)\)] (c1),
};
\end{feynman}
\end{tikzpicture}
\end{center}
\vspace{-0.7cm}
\caption{Majorana mass matrices $m_L$ and $m_R$ at one-loop level. Here, {$Z_3$-TY} charges are given in parenthesis. One clearly finds that this one-loop radiatively violates {$Z_3$-TY} symmetry since $a\otimes a\neq {\mathbbm{1}}$ although each of vertex is invariant under {$Z_3$-TY}.   }
\label{fig:Mass}
\end{figure}

\section{Model setup}
\label{sec:II}

\begin{table}[t!]
\begin{tabular}{|c||c|c|c|c|c||c|c|}\hline\hline  
& ~$L_L$~& ~$\ell_R$~ & ~$N_L$~ & ~${N_R}$~ & ~${X_R}$~ & ~$H$~  & ~{$S$}~    \\\hline\hline 
$SU(2)_L$   & $\bm{2}$  & $\bm{1}$  & $\bm{1}$ & $\bm{1}$ & $\bm{1}$   & $\bm{2}$      & $\bm{1}$  \\\hline 
$U(1)_Y$    & $-\frac12$  & $-1$  & $0$  & $0$    & $0$ & $\frac12$      & $0$    \\\hline
{$Z_3$-TY}   & ${a}$  & $ a $& $ a$ & $a$ & $n $  & $\mathbbm{I}$ & $n$ \\\hline 
 \end{tabular}
\caption{Charge assignments of the fields under the $SU(2)_L \times U(1)_Y$ gauge symmetry and $Z_3$-TY.
}\label{tab:1}
\end{table}

Here, we show our model.

\subsection{Neutral fermions}
In addition to the ISS particles in Eq.~(\ref{eq:iss}), we introduce three right-handed neutral fermions $X_R$ and an inert singlet boson $S\equiv [S_R + i S_I]/\sqrt2$. These two particles are assigned by, $n$ while the other particles except $H$ are assigned by $a$.
The SM Higgs is neutral under {$Z_3$-TY}. All the particle contents and their assignments are listed in Table~\ref{tab:1}.
Under this symmetry, $m_{R,L}$ are forbidden at tree level since ${\mathbbm I}\notin a^2$ and ${\mathbbm I}\notin  (a^*)^2$.
Below we show forbidden renormalizable terms due to  {$Z_3$-TY};
\begin{align}
& {\overline N^C_L} N_L,\quad {\overline N^C_R} N_R,\quad {\overline N^C_L} N_L S,\quad {\overline N^C_R} N_R S, \quad 
{\overline N_L} X_R,\quad {\overline L_L} \tilde H N^C_L , \label{eq:forbid1} \\
& {\overline N^C_R} X_R,\quad {\overline L_L} \tilde H X_R  S^3, \label{eq:forbid2}
\end{align}
where  $\tilde H\equiv i\sigma_2 H^*$, $\sigma_2$ being the second Pauli matrix.
But all the other terms in Eq.~(\ref{eq:iss}) are allowed.
Then, $N_{L,R}$ can have the following interactions which are invariant under {$Z_3$-TY} symmetry %
\begin{align}
f_{L_{\alpha a}} \overline{X_{R_\alpha}} N_{L_a} S 
+
f_{R_{a\alpha}} \overline{N_{R_a}} X^C_{R_\alpha} S
+
M_{X_\alpha}  \overline{X_{R_\alpha}^C} X_{R_\alpha}
-\mu^2_S S^2
+{\rm h.c.}. \label{eq:m_RL}
 \end{align}
 The above terms lead us to generate $m_{R,L}$ at one-loop level as shown in Fig.~\ref{fig:Mass}.
 One clearly finds that the one-loop diagram radiatively violates {$Z_3$-TY} symmetry since $a\otimes a\neq {\mathbbm{1}}$ although each of vertex is invariant under {$Z_3$-TY}.
 As a result, we find the following mass terms:
 \begin{align}
 m_{L_{ab}}&=
\frac1{2(4\pi)^2}
\sum_{\alpha=1}^3  
{f^T_{L_{a\alpha}} M_{X_\alpha} f_{L_{\alpha b}} }
\left[
\frac{m^2_{S_R}}{m^2_{S_R} -M_{X_\alpha}}\ln\left(\frac{m^2_{S_R}}{M_{X_\alpha}^2}\right)
-
\frac{m^2_{S_I}}{m^2_{S_I} - M_{X_\alpha}}\ln\left(\frac{m^2_{S_I}}{M_{X_\alpha}^2}\right)
\right], \\
m_{R_{ab}}&=
\frac1{2(4\pi)^2}
\sum_{\alpha=1}^3  
{f_{R_{a\alpha}} M_{X_\alpha} f^T_{R_{\alpha b}} }
\left[
\frac{m^2_{S_R}}{m^2_{S_R} -M_{X_\alpha}}\ln\left(\frac{m^2_{S_R}}{M_{X_\alpha}^2}\right)
-
\frac{m^2_{S_I}}{m^2_{S_I} - M_{X_\alpha}}\ln\left(\frac{m^2_{S_I}}{M_{X_\alpha}^2}\right)
\right], \label{eq:m_lR-form}
\end{align}
where $m_{S_R}$ and $m_{S_I}$ are respectively the mass eigenvalues of $S_R$ and $S_I$.
Here, we redefine to be $m^2_0\equiv (m_{S_R}^2+m_{S_I}^2)/2$, $\delta m^2\equiv (m_{S_R}^2 - m_{S_I}^2)/2$, and we suppose $\delta m^2\ll M_{X_\alpha}^2$.  Furthermore, $r_\alpha \equiv m^2_0/M_{X_\alpha}^2$.
Then, we rewrite them as follows:
 \begin{align}
 m_{L_{ab}}&\simeq-
\frac{\delta m}{(4\pi)^2}
\sum_{\alpha=1}^3  
 \frac{\delta m}{M_{X_\alpha}}
\frac{f^T_{L_{a\alpha}} f_{L_{\alpha b}}}{(1-r_\alpha)^2} 
\left(
1-r_\alpha+\ln[r_\alpha]
\right), \\
 m_{R_{ab}}&\simeq-
\frac{\delta m}{(4\pi)^2}
\sum_{\alpha=1}^3  
 \frac{\delta m}{M_{X_\alpha}}
\frac{f_{R_{a\alpha}} f^T_{R_{\alpha b}}}{(1-r_\alpha)^2} 
\left(
1-r_\alpha+\ln[r_\alpha]
\right). \label{eq:m_lR-approx}
\end{align}
%
%
%

{
\subsection{Why the \(Z_3\) Tambara–Yamagami selection rule?}
An important feature of the present construction is that the \(Z_3\)-TY fusion rule simultaneously controls the allowed and forbidden interactions required for radiative inverse seesaw masses. In particular, the fusion rule
\[
n\otimes n={\mathbbm I}\oplus a\oplus a^*
\]
allows the singlet channels responsible for $\overline{X_{R}^C} X_{R}$ and $S^2$ terms in Eq. (11).
Furthermore, since
\[
a^{(*)}\otimes n\otimes n
={\mathbbm I} \oplus a\oplus a^*,
\]
the interactions $\overline{X_{R}} N_{L} S $ and $
\overline{N_{R}} X^C_{R} S$ are also allowed. 
At the same time,  as shown in Eqs.~(\ref{eq:forbid1}) and (\ref{eq:forbid2}), the Majorana mass terms, Dirac mass terms, and the cubic scalar interaction of \(S^3\) are forbidden by the TY selection rule.

This simultaneous selection of interactions is essential for our construction. The allowed interactions proportional to $f_{L,R}$ in Eq.~(\ref{eq:m_RL})
generate the Majorana mass parameters \(m_L\) and \(m_R\) radiatively, whereas the absence of the \(S^3\) interaction is important for avoiding the unwanted logarithmic divergence. 

It is worth emphasizing that this operator structure cannot be reproduced by a simple Abelian discrete symmetry with one-dimensional charge assignments. 
To see this, let $q_\phi$ denote the Abelian charge of a field $\phi$. Requiring the Majorana mass term of $X_R$, the $S^2$ term, and the two interactions proportional to $f_{L,R}$ in Eq. (11) to be invariant gives
\[
\begin{aligned}
2q_X&=0,\  2q_S=0,\
-q_X+q_{N_L}+q_S=0,\
- q_{N_R}-q_X+q_S=0.
\end{aligned}
\]
Consequently,
\[
2q_{N_L}=2q_{N_R}=0,
\]
which implies that the Majorana mass terms for \(N_L\) and \(N_R\) are also allowed by the same Abelian symmetry.
This contradicts the requirement that these terms be forbidden at tree level.
Thus, the $Z_3$-TY fusion rule provides a unified realization of the required interaction structure without introducing additional Abelian selection rules.

We do not claim that the resulting low-energy observables uniquely identify a non-invertible symmetry, since similar effective interactions may potentially be obtained from more elaborate conventional symmetry constructions. Rather, the main advantage of the $Z_3$-TY framework is its ability to organize the required operator selection and radiative mass generation within a single non-invertible fusion rule.

}

{
\subsection{Dark matter candidate}
Here, we briefly comment on our DM candidate. 
In addition to the interactions in Eq.~(11), the scalar $S$ can interact with the SM Higgs via the Higgs portal coupling $\lambda_{HS} |H|^2 |S|^2$. 
This term is strictly allowed by the $Z_3$-TY selection rule since $|S|^2$ contains the trivial representation $I$ via the fusion rule $n \otimes n = I \oplus a \oplus a^*$. 
This portal coupling plays a crucial role in the dark matter phenomenology discussed below.
The stability of the DM is assured by the $Z_3$-TY selection rule, making the lightest particle with charge $n$ stable. In our setup, the scalar $S$ serves as the primary DM candidate, as the fermionic $X_R$ lacks sufficient interactions to yield the correct relic density. 

The scalar $S$ interacts with the SM sector primarily via the Higgs portal. As discussed in ref.~\cite{Kanemura:2010sh}, the viable parameter space for such a singlet scalar DM is constrained to two regions: the Higgs resonance region ($m_S \simeq m_h/2$) and the heavy mass region ($m_S \gtrsim \mathcal{O}(1)$ TeV). 

In our numerical analysis (Tables II and III), we focus on the Higgs resonance region, setting $m_S \simeq m_h/2$. In this region, the annihilation cross-section is enhanced by the $s$-channel Higgs exchange, naturally suppressing the relic density to the observed value ($\Omega h^2 \simeq 0.12$) even with a small Higgs-DM coupling. Furthermore, the direct detection constraints (e.g., from XENONnT or LZ) are evaded because the DM-nucleon scattering cross-section is sufficiently suppressed in this narrow resonance window. 
Thus, the benchmark points presented in this work are fully consistent with current cosmological and direct detection bounds, without requiring additional fine-tuning.
}

\subsection{Active neutrinos}
The active neutrino mass matrix is given in Eq~(\ref{eq:iss}), where we rewrite each of the mass matrix as follows:
\begin{align}
M_D &=M_3
\left[\begin{array}{ccc}
\tilde M_1 & 0& 0\\
0 & \tilde M_2 & 0 \\
0 & 0 & 1 \\
\end{array}\right]\equiv M_3 \tilde M_D,\\
m_{D_R} &=\frac{v_H y_{33}}{\sqrt2}
\left[\begin{array}{ccc}
 y_{11} & y_{12} & y_{13}\\
y_{21} & y_{22} & y_{23}\\
y_{31} & y_{32} & 1\\
\end{array}\right]
\equiv \frac{v_H y_{33}}{\sqrt2} \tilde m_{D_R},\\
m_L &\equiv -\frac{\delta m}{(4\pi)^2}\tilde m_L.
\end{align}
 Then, $m_\nu$ is rewrite as 
\begin{align}
m_\nu
&\simeq -\frac{y^2_{33}}{2(4\pi)^2}\frac{v^2_H }{M_3^2}  \delta m 
\left[\tilde m^T_{D_R} (\tilde M_D)^{-1} \tilde m_L \tilde M_D^{-1} \tilde m_{D_R}\right]
\equiv \kappa_\nu\tilde m_\nu. \label{eq:iss_ours}
\end{align}
$m_\nu$ is diagonalized by a unitary matrix $U$, that is,  $D_\nu=U^T m_\nu U\equiv \kappa_\nu U^T \tilde m_\nu U$.
Furthermore $\tilde D_\nu \equiv{\rm diag}[\tilde D_1, \tilde D_2,\tilde D_3] \equiv D_\nu/\kappa_\nu$.
{\it $U$ is observed matrix where charged-lepton mass matrix can be diagonal without loss of generality.}
Then, two  squared mass differences $\Delta m^2_{atm}$ and $\Delta m^2_{sol}$ can be written in terms of our parameters as follows:
\begin{align}
\kappa^2_\nu=\frac{\Delta m^2_{sol}}{\tilde D_2^2-\tilde D_1^2}, \label{eq:kappa}
\end{align}
and
\begin{align}
&{\rm NH}: \Delta m^2_{atm} = \Delta m^2_{sol}\left( \frac{\tilde D_3^2-\tilde D_1^2}{\tilde D_2^2-\tilde D_1^2}\right),\\
&{\rm IH}: \Delta m^2_{atm} = \Delta m^2_{sol}\left( \frac{\tilde D_2^2-\tilde D_3^2}{\tilde D_2^2-\tilde D_1^2}\right).
\end{align}
Sum of neutrino masses; $\sum D_\nu=\kappa_\nu(\tilde D_1  +\tilde  D_2+\tilde  D_3)$, should be less than 120 meV~\cite{Planck:2018vyg} coming from the minimal standard cosmological model with CMB data.
%
The effective mass for neutrinoless double beta decay $\langle m_{ee}\rangle$ is defined as follows
\begin{align}
\langle m_{ee}\rangle= \kappa_\nu \left| \tilde D_{1} c^2_{12} c^2_{13}+ \tilde D_{2} s^2_{12} c^2_{13}e^{i\alpha}+ \tilde D_{3} s^2_{13}e^{i(\beta-2\delta_{CP})} \right|,
\end{align}
where $s_{12,23,13} (c_{12,23,13})$, which are short-hand notations $\sin\theta_{12,23,13} (\cos\theta_{12,23,13})$, are neutrino mixing of $U$, $\delta_{CP}$ is Dirac phase, $P\equiv {\rm diag}[1,e^{i\alpha/2},e^{i\beta/2}]$ are Majorana phases, $P$ being the Majorana phase matrix. 
The upper bound on $\langle m_{ee}\rangle$ is given by the current KamLAND-Zen data measured~\cite{KamLAND-Zen:2024eml}
that tells us $ \langle m_{ee}\rangle  <(28-122)$ meV at 90 \% confidence level (CL).

{
\subsection{Stability of the Mass Hierarchy and Absence of Tree-level $m_{D_L}$}
Before proceeding to the numerical analysis, we provide an argument that the mass term $m_{DL}$ is strictly forbidden at all loop orders, addressing the concern regarding the stability of the mass hierarchy. 
The mass term $m_{D_L}$ corresponds to the gauge-invariant operator $\mathcal{O}_{D_L} \equiv \overline{L_L} \tilde{H} N_L$ (and its Hermitian conjugate), where $\tilde{H} \equiv i\sigma_2 H^*$. 
Under the $Z_3$-TY selection rule, the charge of this operator is given by $a \otimes I \otimes a = a^*$. 
Since the Lagrangian must be invariant (charge $I$), the radiative generation of $\mathcal{O}_{D_L}$ requires a source with charge $a$ to form a singlet ($a^* \otimes a = I$).

At the tree level, such a source would be the vacuum expectation value (VEV) of a scalar field carrying charge $a$. 
However, as listed in Table~I, all fields with charge $a$ ($L_L, N_L, N_R$) are fermions, and no scalar field in the model spectrum possesses charge $a$. 
Since fermion fields cannot acquire VEVs without breaking Lorentz invariance, $m_{D_L}$ is strictly forbidden at the tree level.

One might wonder if $m_{D_L}$ could be generated at the loop level, similar to $m_{L,R}$. 
The generation of $m_{L,R}$ is allowed because the external fields $N_{L,R}$ are $SU(2)_L$ singlets, and the internal particles in the loop ($X_R$ and $S$ with charge $n$) can effectively provide the necessary charge $a$ via the non-invertible fusion rule $n \otimes n = I \oplus a \oplus a^*$, without conflicting with the gauge symmetry.

In contrast, the operator $\mathcal{O}_{D_L}$ involves the $SU(2)_L$ doublet field $L_L$. 
To construct a loop diagram that generates $\mathcal{O}_{D_L}$ while preserving both the $Z_3$-TY symmetry and the $SU(2)_L$ gauge symmetry, the loop must contain an internal particle that simultaneously carries $SU(2)_L$ doublet quantum numbers and $Z_3$-TY charge $a$ (to screen the $a^*$ charge of the operator). 
However, as shown in Table~I, the only $SU(2)_L$ doublet fields in the model are $L_L$ (charge $a$) and $H$ (charge $I$). 
Since $L_L$ is an external fermion and cannot run inside the loop as a scalar propagator, and the Higgs $H$ carries charge $I$ (not $a$), there is no internal scalar field that can mediate the required $SU(2)_L$ doublet structure while providing the charge $a$. 
Consequently, it is topologically and group-theoretically impossible to construct any loop diagram that satisfies both the gauge invariance and the $Z_3$-TY selection rule for $\mathcal{O}_{D_L}$. 
This guarantees that $m_{D_L} = 0$ exactly to all orders in perturbation theory, ensuring the robustness of the hierarchy $m_{D_L} \ll m_{L,R}$.
}

 \section{Numerical results}
 \label{sect.3}
 Here, we perform our numerical analysis and show benchmark points for NH and IH with the minimum chi square.
 Before that, we simplify $m_{D_R}$ to be the diagonal mass matrix; we set $y_{21}=y_{31}=y_{32}=y_{12}=y_{13}=y_{23}\sim 0$, and $\tilde m_{D_R}\sim {\rm diag}[y_{11},y_{22},1]$. $f_L$ is three by three complex dimensionless parameters, therefore we suppose that observed mixing angles and phases originates from $f_L$.
Since we simplify $m_{D_R}$, Max[$|\tilde M_D^{-1} \tilde m_{D_R}|$]$=1$. 
That is, the constraint of $|F|$ is found  as follows:
\begin{align}
& |F|=\left(\frac{v_H y_{33}}{\sqrt2 M_3}\right) |\tilde M_D^{-1} \tilde m_{D_R}| \lesssim 4.9\times 10^{-3},\\
&q_{min}\equiv \left(\frac{v_H y_{33}}{\sqrt2 M_3}\right)_{min}\lesssim \frac{4.9\times 10^{-3}}{ |\tilde M_D^{-1} \tilde m_{D_R}|_{max}}
=  4.9\times 10^{-3}.\label{eq:q}
\end{align}
On the other hand, since $\kappa_\nu$ is experimentally determined in Eq.~(\ref{eq:kappa}),
$q_{min}$ in Eq.~(\ref{eq:q}) is rewritten in terms of $\kappa_\nu$ and $\delta m$ and, we find the following condition
\begin{align}
(4\pi) \sqrt{\frac{\kappa_\nu}{\delta m}} \lesssim  4.9\times 10^{-3}.
\end{align}
In addition to the above constraints, we adopt Nufit 6.0~\cite{Esteban:2024eli} as experimental observables and remind it $m_0\simeq m_S=m_h/2$.
We estimate $\chi^2$ by applying the following formula
\begin{align}
\chi^2 = \sum_{i}  \left( \frac{O_i^{\rm obs} - O_i^{\rm th}}{\delta O_i^{\rm exp}} \right)^2, \label{eq:chi-square}
\end{align}
where $O_i^{\rm obs (th)}$ is the observed (theoretically obtained) value of the corresponding observables and $\delta O_i^{\rm exp}$ indicates the experimental error at $1\sigma$. Here, we adopt five reliable observables; $s_{12},s_{23},s_{13},\Delta m^2_{sol},\ \Delta m^2_{atm}$, assuming all are Gaussian.
{It implies that 1$\sigma$ confidential level($\sim$68.27\%) corresponds to $\chi^2\simeq5.89$,
and 3$\sigma$ confidential level($\sim$99.73\%) corresponds to $\chi^2\simeq18.21$. }

{
\subsection{Best fit values}
At first, we show the best fit values (BFs) for NH and IH, instead of global analysis.
In case of {\it NH} case, we obtain the best fit value of $\chi^2$ is 0.0593 and show our input values and observables in Table~\ref{BP-nh}.
In case of {\it IH} case, we obtain the best fit value of $\chi^2$ is 0.0673 and show our input values and observables in Table~\ref{BP-ih}.

Comparing the results for NH and IH, we find a significant difference in the effective mass \(\langle m_{ee}\rangle\): \(\langle m_{ee}\rangle=2.00\) meV for NH and \(47.6\) meV for IH. While the NH best-fit value is below the projected sensitivities of upcoming neutrinoless double beta decay experiments, the IH best-fit value is within the sensitivity reach of next-generation experiments such as LEGEND-1000~\cite{LEGEND:2021bnm}, nEXO~\cite{nEXO:2021ujk}, and CUPID~\cite{CUPID:2025faj}, whose projected sensitivities are typically at the level of several to tens of meV, depending on the nuclear matrix element calculations. Therefore, our IH benchmark point could be tested by these experiments, provided that the standard light-Majorana-neutrino exchange mechanism dominates the decay process.\\

 \begin{table}[h!]
 		\centering
 	\begin{tabular}{|c|c|} \hline 
 		\rule[14pt]{0pt}{0pt}	
 		$\delta m$& $7.66$ GeV  \\ 
		\rule[14pt]{0pt}{0pt}	
 		$m_S$& $62.5$ GeV  \\ 
 		\rule[14pt]{0pt}{0pt}
		$[M_{X_1},M_{X_2},M_{X_3}]$& $[862,\ 2.86\times10^4,\ 5.63\times10^4]$ GeV  \\ 
 		\rule[14pt]{0pt}{0pt}
 		$[\tilde M_{1},\tilde M_{2}]$ &$[0.0168,\ 0.143]$ \\
 		\rule[14pt]{0pt}{0pt}
 		$[y_{11},\ y_{22}]$ &$[0.000848,\ 0.100]$ \\
 		\rule[14pt]{0pt}{0pt}
 		$ f_L \times10^2$ & $
		\left[\begin{array}{ccc}
 -6.83 + 0.621 i & 4.29 + 2.88 i & 1.81 + 1.99 i \\
13.8 + 23.9 i & 1.32 - 3.30 i & 0.0398 + 0.00386 i \\
4.92 - 82.8 i & 3.14 - 15.3 i & 0.0310 - 0.0355 i \\
\end{array}\right]$  \\
 		\rule[14pt]{0pt}{0pt}
         	$|F|_{max}$ & $0.00382$   \\
 		\rule[14pt]{0pt}{0pt}
 	$\kappa_\nu$ & $ 7.08\times 10^{-7}$  \\
 		\rule[14pt]{0pt}{0pt}
	$[\sin\theta_{12},\ \sin\theta_{23},\ \sin\theta_{13}]$ & $ [0.550, 0.758, 0.149]$  \\
 		\rule[14pt]{0pt}{0pt}
 	$[\Delta m^2_{sol},\ \Delta m^2_{atm}] $ & $[7.39\times 10^{-5},\ 2.51\times 10^{-3}]{\rm eV^2}$  \\
 		\rule[14pt]{0pt}{0pt}
 		$[\delta_{CP},\alpha,\beta]$ & $[141^\circ, 343 ^\circ, 109 ^\circ]$ \\
 		\rule[14pt]{0pt}{0pt}
	$[\langle m_{ee}\rangle,\ \sum D_\nu]$ & $ [2.00,\ 59.4]{\rm meV}$  \\
 		\rule[14pt]{0pt}{0pt}
 		$\chi^2$ & $0.0593$ 	\\
 		\hline
 	\end{tabular}
 	\caption{Best fit of $\chi^2$ in case of NH.	 
 	}
 	\label{BP-nh}
 \end{table}
%

 \begin{table}[h!]
 		\centering
 	\begin{tabular}{|c|c|} \hline 
 		\rule[14pt]{0pt}{0pt}	
 		$\delta m$& $5.48$ GeV  \\ 
		\rule[14pt]{0pt}{0pt}	
 		$m_S$& $62.5$ GeV  \\ 
 		\rule[14pt]{0pt}{0pt}
		$[M_{X_1},M_{X_2},M_{X_3}]$& $[138,\ 9.24\times10^3,\ 1.63\times10^4]$ GeV  \\ 
 		\rule[14pt]{0pt}{0pt}
 		$[\tilde M_{1},\tilde M_{2}]$ &$[0.00203,\ 0.0685]$ \\
 		\rule[14pt]{0pt}{0pt}
 		$[y_{11},\ y_{22}]$ &$[0.00540,\ 0.505]$ \\
 		\rule[14pt]{0pt}{0pt}
 		$ f_L \times10^2$ & $
		\left[\begin{array}{ccc}
-0.00236 - 0.0278 i & -21.7 - 79.6 i & 0.00106 + 0.00142 i \\
6.52 - 0.529  i & -0.00651 + 0.00339  i & 0.00509 + 0.00146  i \\
22.0 + 51.7 i & 47.4 - 18.2  i & 0.0713 - 1.31 i \\
\end{array}\right]$  \\
 		\rule[14pt]{0pt}{0pt}
         	$|F|_{max}$ & $0.00171$   \\
 		\rule[14pt]{0pt}{0pt}
 	$\kappa_\nu$ & $ 1.86\times 10^{-9}$  \\
 		\rule[14pt]{0pt}{0pt}
	$[\sin\theta_{12},\ \sin\theta_{23},\ \sin\theta_{13}]$ & $ [0.554, 0.759, 0.149]$  \\
 		\rule[14pt]{0pt}{0pt}
 	$[\Delta m^2_{sol},\ \Delta m^2_{atm}] $ & $[7.51\times 10^{-5},\ 2.50\times 10^{-3}]{\rm eV^2}$  \\
 		\rule[14pt]{0pt}{0pt}
 		$[\delta_{CP},\alpha,\beta]$ & $[92.0^\circ, 21.3^\circ, 323^\circ]$ \\
 		\rule[14pt]{0pt}{0pt}
	$[\langle m_{ee}\rangle,\ \sum D_\nu]$ & $ [47.6,\ 99.2]{\rm meV}$  \\
 		\rule[14pt]{0pt}{0pt}
 		$\chi^2$ & $0.0673$ 	\\
 		\hline
 	\end{tabular}
 	\caption{Best fit of $\chi^2$ in case of IH.	 
 	}
 	\label{BP-ih}
 \end{table}

\subsection{Local correlations nearby BFs}
As can be seen from the structure of our model, it involves a large number of independent input parameters. A comprehensive exploration of the full parameter space is therefore highly nontrivial, and any global correlations inferred from a restricted numerical scan could strongly depend on the chosen scan ranges. Instead, we focus on the parameter space in the vicinity of the best-fit point to illustrate the local behavior of the model and the correlations among relevant observables. We emphasize that these correlations characterize only the neighborhood of the best-fit point and should not be interpreted as global features or generic predictions of the model.
\\

To investigate the local behavior of the model, we perform a numerical scan in which each input parameter is varied within approximately \(\pm 10\%\) of its best-fit value for each hierarchy.

\subsubsection{NH case}
\begin{figure}[tb]\begin{center}
\includegraphics[width=80mm]{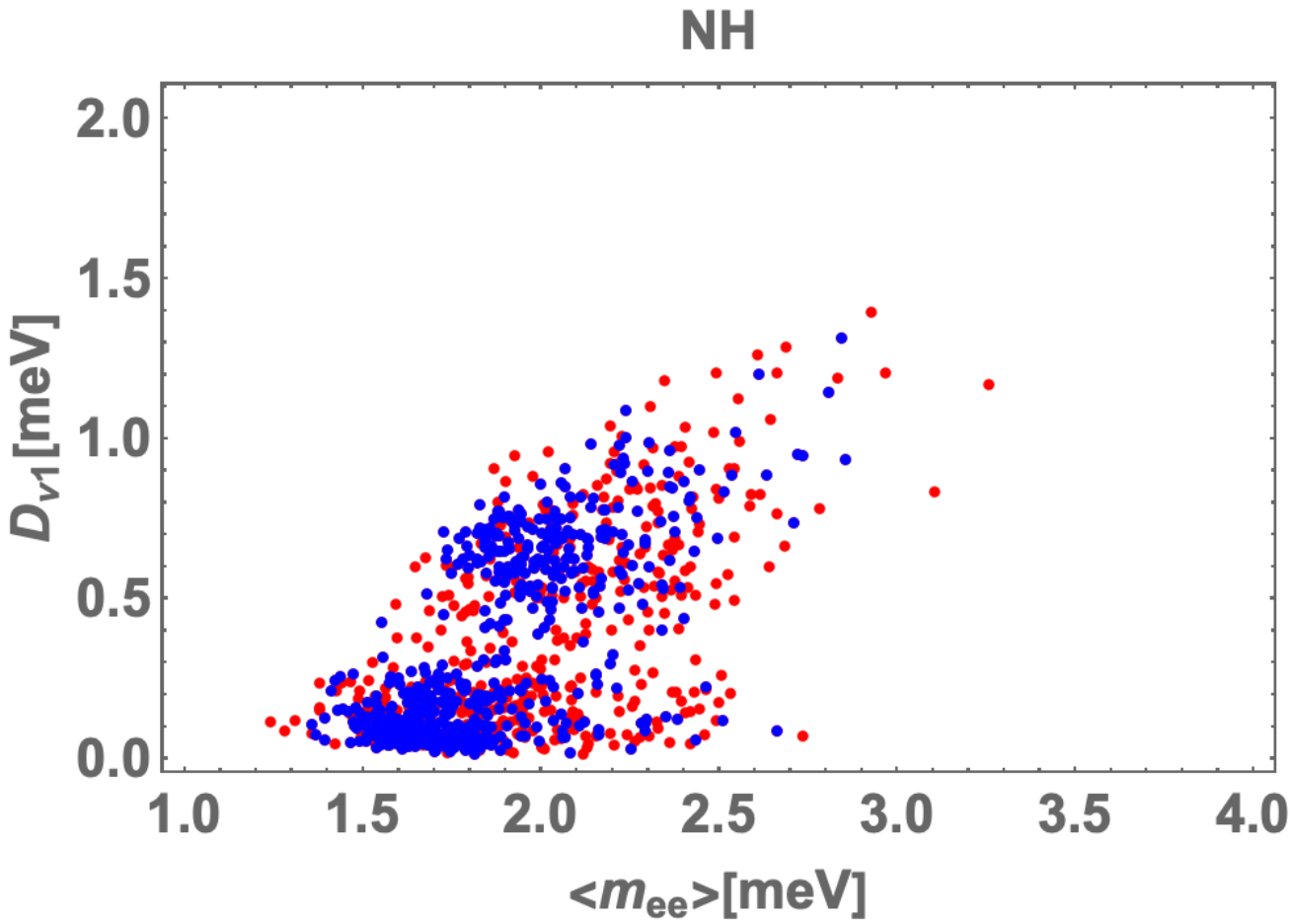}\quad
\includegraphics[width=80mm]{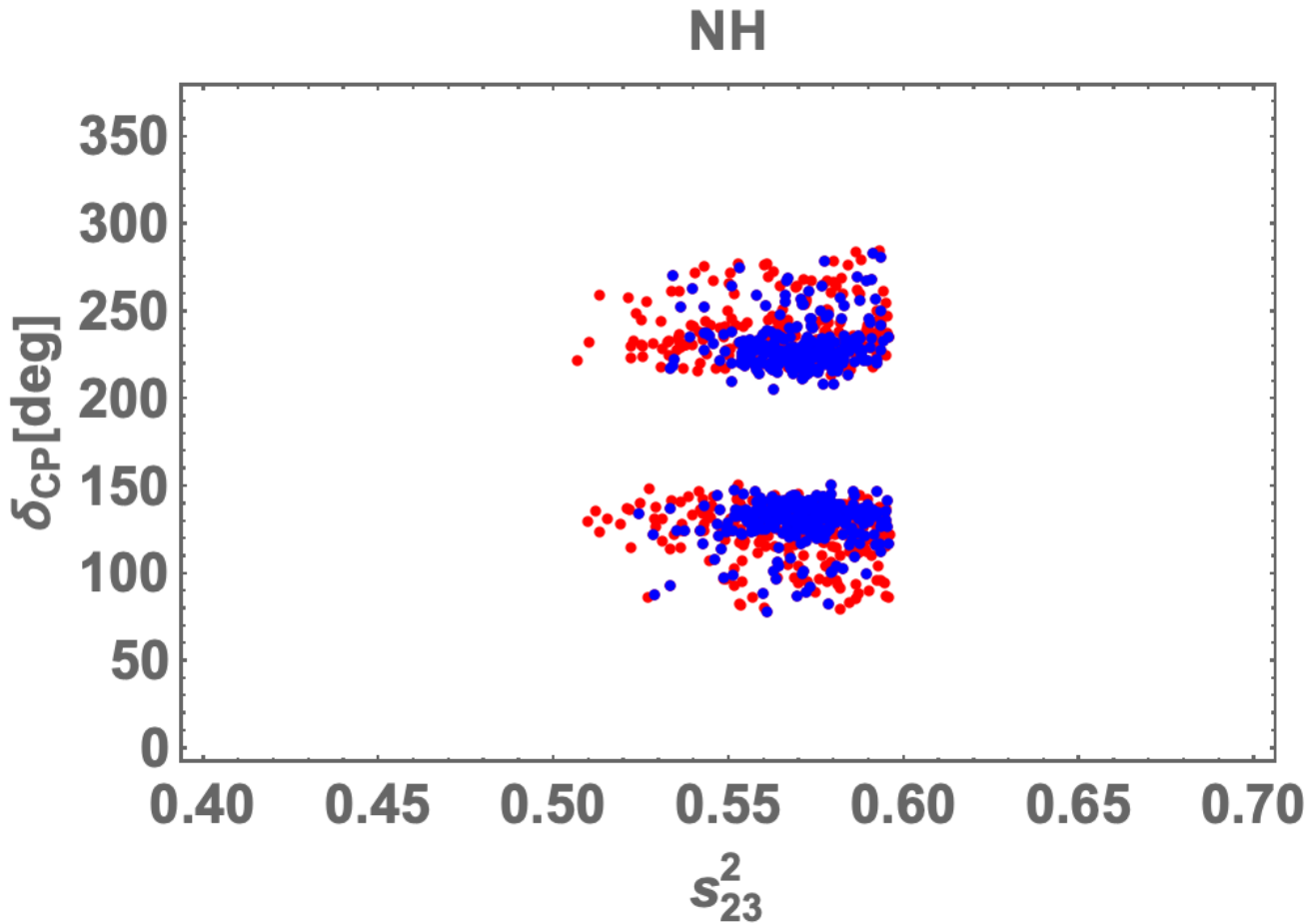}
\caption{The local allowed regions of the lightest neutrino mass $D_{\nu_1}$ in terms of $\langle m_{ee}\rangle$ (left-panel) and $\delta_{\rm CP}$ in terms of $s^2_{23}$ (right-panel) in case of NH.}   
\label{fig:nh}
\end{center}\end{figure}
%
Fig.~\ref{fig:nh} shows the local allowed regions of the lightest neutrino mass \(D_{\nu_1}\) versus \(\langle m_{ee}\rangle\) (left panel) and \(\delta_{\rm CP}\) versus \(s^2_{23}\) (right panel). The blue and red points represent the parameter sets within the \(1\sigma\) and \(3\sigma\) ranges, respectively. We find that \(D_{\nu_1}\) is restricted to values below 1.5 meV, while \(\langle m_{ee}\rangle\) is below 3.5 meV, which is well below the most stringent upper bound of 28 meV from KamLAND-Zen. The Dirac CP phase \(\delta_{\rm CP}\) favors two regions, approximately \(60^\circ\)–\(130^\circ\) and \(210^\circ\)–\(270^\circ\), whereas \(\sin^2\theta_{23}\) spans nearly the entire range allowed at the \(3\sigma\) level by current experimental data.


\subsubsection{IH case}
\begin{figure}[tb]\begin{center}
\includegraphics[width=80mm]{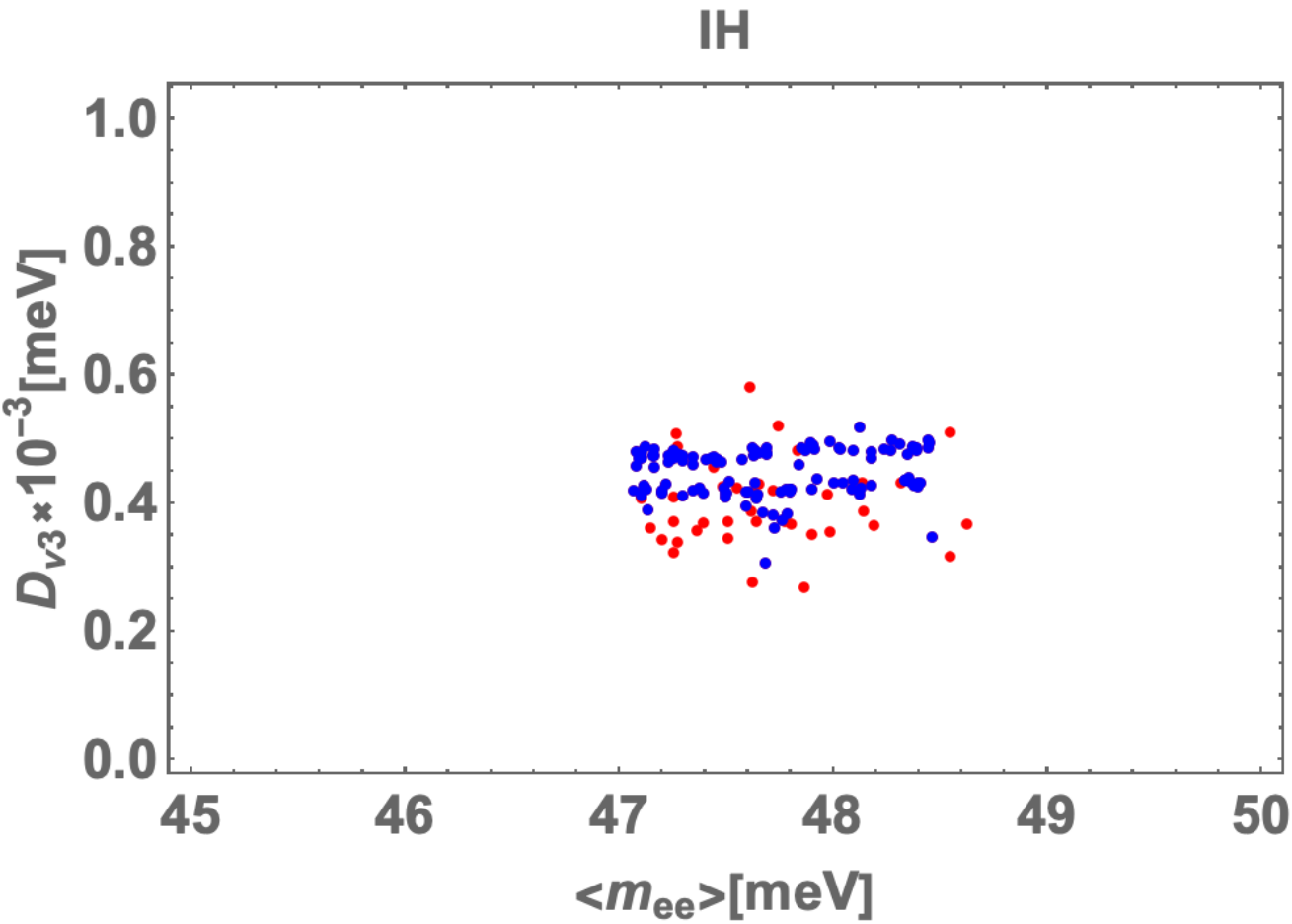}\quad
\includegraphics[width=80mm]{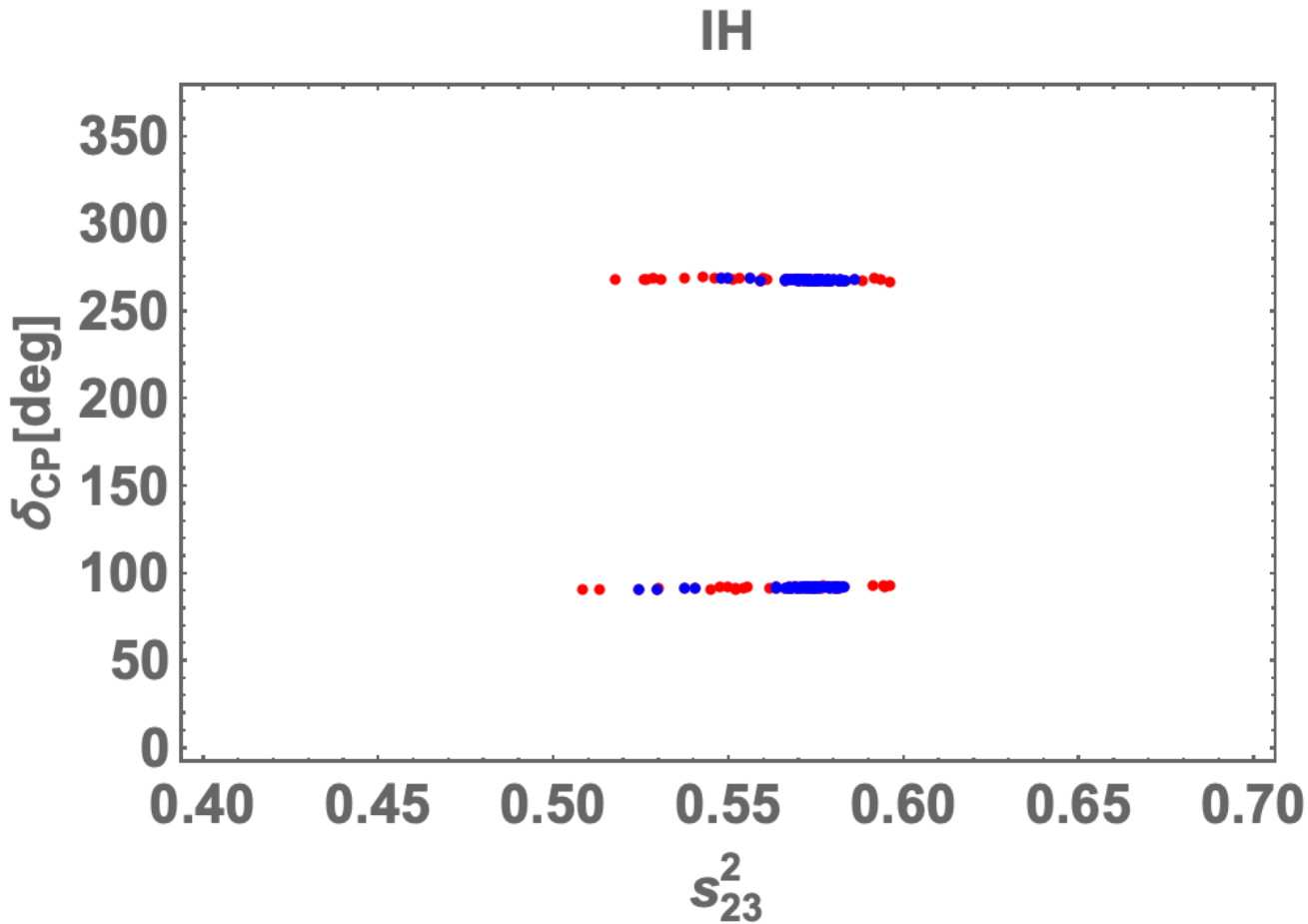}
\caption{The local allowed regions of the lightest neutrino mass $D_{\nu_3}$ in terms of $\langle m_{ee}\rangle$ (left-panel) and $\delta_{\rm CP}$ in terms of $s^2_{23}$ (right-panel) in case of IH.}   
\label{fig:ih}
\end{center}\end{figure}
%
Fig.~\ref{fig:ih} shows the local allowed regions of the lightest neutrino mass \(D_{\nu_3}\) versus \(\langle m_{ee}\rangle\) (left panel) and \(\delta_{\rm CP}\) versus \(\sin^2\theta_{23}\) (right panel). The color coding is the same as in the NH case. We find that \(D_{\nu_3}\) is required to be nearly zero, while \(\langle m_{ee}\rangle\) is confined to the range of 47–49 meV, which lies within the upper-bound range of 28–122 meV obtained by KamLAND-Zen. The Dirac CP phase \(\delta_{\rm CP}\) is localized around \(90^\circ\) and \(270^\circ\), whereas \(s^2_{23}\) spans nearly the entire range allowed at the \(3\sigma\) level by current experimental data.


}

\section{Summary and discussion}
In this work, we have demonstrated that the $Z_3$ Tambara-Yamagami selection rule provides a  natural realization of the mass hierarchies required in the inverse seesaw framework. This symmetry forbids Majorana mass terms for $N_R$ and $N_L$ at the tree level, while these terms are generated at the one-loop level through the radiative symmetry breaking of the selection rule.
 This mechanism not only explains the small neutrino masses without fine-tuning but also naturally incorporates a viable dark matter candidate, which contributes to the generation of the suppressed mass terms. Furthermore, by performing a numerical analysis based on Nufit 6.0 data, we have obtained best-fit results for both normal and inverted neutrino mass hierarchies, showing consistency with current experimental observations. Our findings highlight the effectiveness of the $Z_3$-TY symmetry in addressing neutrino phenomenology and its potential connection to dark matter physics.

\begin{acknowledgments}
 HO is supported by Zhongyuan Talent (Talent Recruitment Series) Foreign Experts Project. 
\end{acknowledgments}

\bibliography{ISS_Z3TY}

\end{document}